\documentclass{edm_article}

\usepackage{natbib}
\usepackage{subcaption}  
\usepackage{listings}
\usepackage{xcolor}
\usepackage{multirow}
\usepackage{booktabs}
\usepackage{url}

\begin{document}

\title{What Changes When the Interlocutor Is an AI? Interactional Fluency and Linguistic Uptake in L2 Spoken Dialogue}

\numberofauthors{6}

\author{
% 1st author
\alignauthor
Russell Scheinberg\\
       \affaddr{Portland State University}\\
       \email{rschein2@pdx.edu}
% 2nd author
\alignauthor
Ameeta Agrawal\\
       \affaddr{Portland State University}\\
       \email{ameeta@pdx.edu}
% 3rd author
\alignauthor
Tetyana Sydorenko\\
       \affaddr{Portland State University}\\
       \email{tsydorenko@pdx.edu}
\and
% 4th author
\alignauthor
Kalab Kahsay\\
       \affaddr{Portland State University}\\
       \email{kalabk@pdx.edu}
% 5th author
\alignauthor
Nina Vyatkina\\
       \affaddr{University of Kansas}\\
       \email{vyatkina@ku.edu}
% 6th author
\alignauthor
Griet Boone\\
       \affaddr{University of Antwerp}\\
       \email{griet.boone@uantwerpen.be}
}

\maketitle

\begin{abstract}
Voice-based AI systems are increasingly used for L2 speaking practice, but evaluations rarely characterize the interactional processes they create. We analyze 78 university learners of German across four sites completing a counterbalanced spot-the-difference task with both a human peer and a real-time AI partner. From diarized ASR transcripts, we extract measures of interactional fluency, linguistic uptake, and learner experience. Human dialogue was faster and more balanced, with many short turns; AI dialogue resembled supported monologue, with fewer, longer turns, reduced learner floor share, and greater within-turn fluency. The AI's verbose, syntactically regular input was associated with greater short-term uptake and stronger syntactic priming after controlling for input volume. Attitudes toward AI improved after the task, and satisfaction was predicted by production fluency rather than uptake. The results show complementary affordances for AI and human dialogue in L2 practice.
\end{abstract}
% \begin{abstract}
% Voice-based AI systems are increasingly used for second language speaking practice, yet evaluations typically emphasize learning outcomes rather than interactional processes. We introduce a scalable, transcript-based framework for analyzing AI-mediated spoken interaction across interactional fluency, linguistic uptake, and learner experience. Seventy-eight university learners of German at four sites completed a spot-the-difference task under four counterbalanced, within-speaker conditions: two monologues and two dialogues (with a human peer and with a real-time AI partner). Using diarized automatic transcripts, we extract temporal and linguistic measures without manual linguistic coding. Human dialogue is rapid and interactive, with short turns, fast responses, and balanced floor time. AI dialogue resembles a supported monologue, with longer turns, slower responses, and reduced learner floor share, but greater within-turn fluency. 
% The AI's verbosity and syntactic regularity are associated with greater short-term linguistic uptake, with learners showing stronger short-term syntactic alignment even after controlling for input volume. Learner attitudes toward AI improve after the task, with satisfaction predicted by production fluency rather than uptake. These results suggest that human and AI dialogue scaffold L2 production through distinct interactional mechanisms, with implications for the design of voice-based language learning systems.

% \end{abstract}


\section{Introduction}

Voice-based conversational AI is increasingly positioned as a scalable speaking partner for second language (L2) learners. Yet these systems do not simply provide more practice, they change the \emph{nature} of the interaction \cite{bibauw2019_dialoguebasedcall,bibauw2022_metaanalysis, porcheron2018voiceinterfaces, fischer2019progressivity, mayor2025can}. AI-based interlocutors, especially using large language models (LLMs), produce longer turns, respond at atypical moments, and lack the backchannels and floor-negotiation strategies of human partners
\cite{mayor2025can,porcheron2018voiceinterfaces,fischer2019progressivity,putz2024performance,choi2026turntakingchatgpt}.
If interaction structure shapes learning opportunities, then characterizing \emph{how} AI-mediated dialogue differs from human dialogue is essential for informed system design and deployment.

% maybe add: skantze2021_turntakingreview,lin2022_duplex
% --- GAP: CALL literature focuses on outcomes, not process ---
Most evaluations of dialogue-based computer-assisted language learning (CALL) focus on outcome measures: pre/post speaking scores, perceived confidence, or learner satisfaction. Meta-analytic evidence confirms positive effects overall but highlights substantial heterogeneity across systems, tasks, and evaluation criteria \citep{bibauw2022_metaanalysis, huang2022_chatbots_review, du2024_chatbots_speaking_review}. What remains underexplored is the \emph{interactional process} itself—how learners and systems jointly construct spoken interaction during practice \citep{LAI2024100291}.

% --- INTERACTIONAL FLUENCY: one clean definition ---
We address this gap through three complementary lenses. The first is \emph{interactional fluency}: measurable properties of conversational coordination---response latency, turn-length distributions, floor sharing, and within-turn pausing---that reflect how speakers manage participation in real-time dialogue. While the distinction between monologic and dialogic fluency is well established in SLA research \citep{tavakoli2016_dialogic_fluency}, less is known about how these dynamics shift when the interlocutor is a machine.

% --- UPTAKE: brief framing ---
The second is \emph{linguistic uptake}. Conversational alignment theory holds that interlocutors converge on lexical and syntactic forms during interaction \citep{pickering2004interactive}, a process proposed as a mechanism supporting L2 learning \citep{Costa01062008, McDonough_2006}. When partners differ fundamentally in how they generate language, it is an open question whether they afford similar uptake opportunities. We operationalize this as direction-sensitive lexical and morphosyntactic uptake: the extent to which a learner reuses words or constructions introduced by the interlocutor in the preceding context.

% --- ATTITUDES: motivate RQ3 ---
The third is \emph{learner experience}. If AI and human dialogue create different interactional conditions, learners may perceive these differences, and the behavioral features of the interaction (e.g., production fluency, floor balance) may predict satisfaction independently of what learners linguistically adopt.

% --- THE STUDY ---

%To investigate these questions, we conducted a controlled, multi-site study with second- and third-year L2 German learners ($N = 78$) at four universities in the United States and Europe. Participants completed a spot-the-difference task under four within-speaker conditions: two monologic scene descriptions and two dialogues (one with a human peer, one with a real-time AI partner). All measures are extracted from diarized, timestamped ASR transcripts without manual annotation.

\noindent These three lenses motivate the following research questions:

\begin{description}
    \item[RQ1:] How do \textbf{interactional fluency} measures differ between {human--human} and {human--AI} dialogue?
    \item[RQ2:] Do learners differ in \textbf{lexical and morphosyntactic uptake} from a human partner versus an AI-interlocutor, and do these differences vary across linguistic levels?
    \item[RQ3:] How do learner \textbf{attitudes toward AI} change after direct experience, and what behavioral features of the interaction predict satisfaction?
\end{description}

To investigate these questions, we conducted a controlled, multi-site study with second- and third-year L2 German learners ($N = 78$) at four universities in the United States and Europe. Participants completed a spot-the-difference task under four within-speaker conditions: two monologic scene descriptions and two dialogues (one with a human peer, one with a real-time AI partner). All measures are extracted from diarized, timestamped ASR transcripts without manual coding.
Results reveal two distinct interaction modes: human dialogue is rapid and balanced, whereas AI dialogue resembles a supported monologue, with fewer but longer turns, slower responses, reduced learner floor share, and greater within-turn fluency. AI verbosity and syntactic regularity afford more uptake opportunities, and learners show stronger short-term syntactic priming with the AI partner even after controlling for input volume. Attitudes toward AI improve after the task, with satisfaction tied to production fluency rather than linguistic uptake.

% Results reveal two qualitatively different interaction modes. Human dialogue is rapid and interactive: frequent short turns, fast responses, and balanced floor time. AI dialogue resembles a supported monologue: fewer but longer turns, slower responses, a reduced learner floor share, but more fluent speech within turns. The greater verbosity and syntactic regularity observed in AI dialogue afford more opportunities for linguistic uptake, but even controlling for amount of speech produced by AI and humans, learners show stronger short-term syntactic priming with the AI partner. These distinct interaction profiles also shape learner experience: attitudes toward AI improve after the task, with satisfaction tied to production fluency rather than linguistic uptake.

% --- CONTRIBUTIONS ---
\noindent This paper makes three contributions: (1)~we operationalize interactional fluency and direction-sensitive uptake for spoken AI practice using metrics extractable from diarized ASR output without manual linguistic annotation; (2)~we leverage a within-speaker design across four instructional sites to isolate how the same learner's interaction changes with a human versus AI partner; and (3)~we provide a reusable analysis pipeline for rapid, scalable evaluation of conversational AI in educational settings. The remainder of the paper describes the study design (\S\ref{sec:study}), methods and results for interactional fluency (\S\ref{sec:temporal-dynamics}), uptake (\S\ref{sec:linguistic-uptake}), and survey analyses (\S\ref{sec:results}), followed by discussion and design implications (\S\ref{sec:discussion}).

\section{Related Work}
\label{sec:previous_work}
Prior work establishes the promise of conversational practice for language learning, but also suggests that who (or what) the interlocutor is can reshape the interaction itself.

\paragraph{Conversational practice and transcript-based measurement}
Dialogue-based computer-assisted language learning (CALL) has a long history motivated by the role of interaction in second language development. Meta-analytic evidence suggests that dialogue-based CALL systems can yield positive learning outcomes overall, while also reporting substantial heterogeneity across system types, tasks, and evaluation measures \citep{bibauw2022_metaanalysis}. Recent reviews of conversational agents for language learning echo both pedagogical promise and persistent concerns about reliability, pedagogical alignment, and evaluation practices \citep{huang2022_chatbots_review, du2024_chatbots_speaking_review}. As voice-based LLM systems become plausible speaking partners, a central question is not only whether learners improve, but what kinds of interaction these systems create and how those interactional properties shape practice opportunities \citep{LAI2024100291}.

A parallel development in learning analytics is the growing treatment of  dialogue as analyzable behavioral trace data. This shift enables process-level evaluation of learning-support interactions, but it also raises a practical challenge for spoken settings: extracting reliable measures without extensive manual annotation. Recent work demonstrates that transcript-based pipelines can extract educationally meaningful structure from naturalistic collaborative speech, enabling scalable, fine-grained analysis \citep{2024.EDM-long-papers.14}. This line of work motivates our use of diarized, timestamped transcripts to quantify interaction dynamics and linguistic behavior across interlocutor conditions and sites.

\paragraph{Interactional fluency and turn-taking with AI partners}

Within second language acquisition (SLA) research, fluency is often decomposed into speed, breakdown, and repair, and dialogic fluency work argues that monologic temporal measures can miss key interactional phenomena in conversation \citep{tavakoli2016_dialogic_fluency}. Overlap, interruptions, and between-turn gaps can materially affect fluency assessments, motivating measurement at the level of turns and turn transitions. Pauses, repetitions, and stalling may likewise function as interactional resources rather than purely as disfluency \citep{peltonen2017_temporal_fluency}. These perspectives build on classic accounts of turn-taking norms that describe strong pressure to minimize gaps and avoid overlap, with response latency as a central coordination variable \citep{sacks1974_turntaking, stivers2009_turntaking, levinson2015_timing}.

\begin{figure*}[t!]
\centering
\includegraphics[trim={0 2.5cm 0 0}, width=1\textwidth]{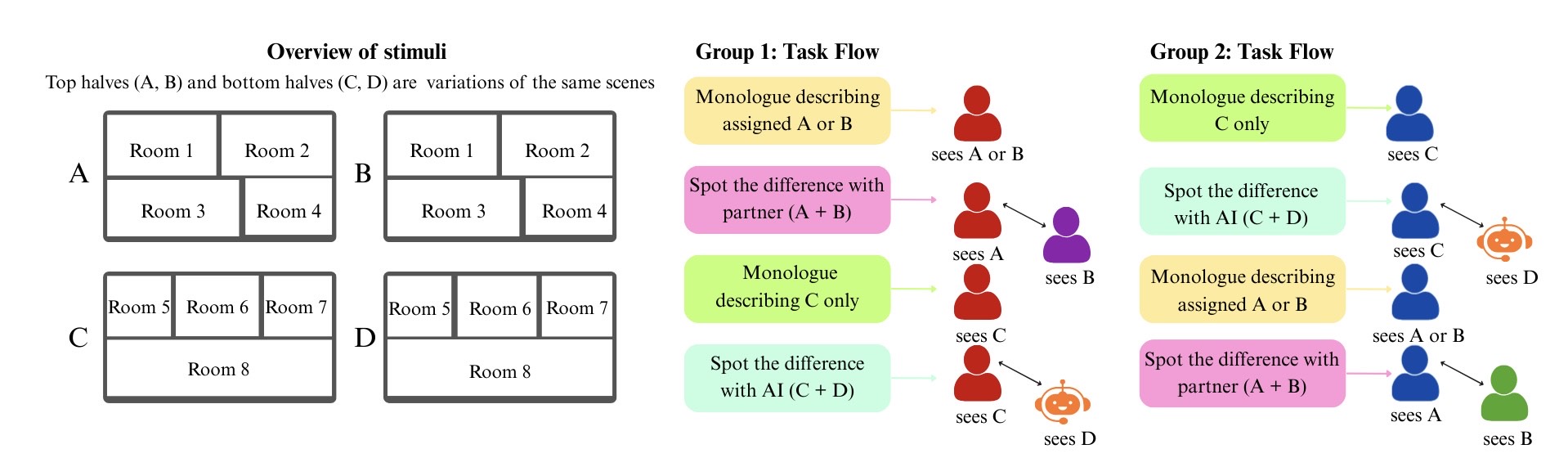}

\caption{Study design. \textbf{Left:} Schema of the two haunted-house scenes, each with four rooms; top halves (A, B) and bottom halves (C, D) are slight variations of the same scenes. In the human condition each partner sees a different version
(A or B), while in the AI condition the participant describes C and the AI is provided a description of D. \textbf{Right:} Task flow for the two counterbalanced groups. Each participant completes four tasks: two monologues and two dialogues (one with a human partner, one with the AI). Group~1 completes the human block first; Group~2 completes the AI block first.}
\Description{Study design diagram showing two haunted-house image sets and the counterbalanced task flow. In the human condition, two partners view different versions of the same scene; in the AI condition, the learner views one image and the AI receives a text description of the paired image.}
\label{fig:experiment-flow}
\end{figure*}

When one interlocutor is a machine, these norms can shift systematically. Studies of voice interfaces document altered turn design and recurring breakdown and repair patterns \citep{porcheron2018voiceinterfaces, luger2016badpa}, and users may perform additional interactional work to maintain progressivity when systems are slow or insufficiently contingent \citep{fischer2019progressivity}. With LLM-based agents, system contributions become more content-rich, yet interface timing and modality still constrain turn-taking and repair; users often scaffold the interaction through reformulations and related strategies \citep{putz2024performance}. Emerging work in language learning contexts suggests that L2 learners’ turn-taking practices may shift when interacting with conversational AI \citep{choi2026turntakingchatgpt}. Interaction structure is also increasingly treated as consequential for learning processes, with dialog acts in AI-supported peer tutoring linked to learning rates \citep{2024.EDM-long-papers.10}. Together, these strands motivate our analysis of interactional fluency as a way to characterize how human--AI dialogue differs from human--human dialogue in the interactional conditions it affords.

\paragraph{Lexical and morphosyntactic uptake as alignment in dialogue}
Psycholinguistic theories of alignment propose that interlocutors tend to converge on lexical and syntactic representations through priming mechanisms, supporting efficient coordination \citep{pickering2004interactive}. Related work on lexical entrainment and conceptual pacts shows how partners establish shared referring expressions as part of building common ground \citep{brennan1996conceptual}. In SLA, alignment has been proposed as a mechanism that can support learning, since repeated exposure to and reuse of forms in interaction may increase accessibility and entrenchment \citep{Costa01062008, McDonough_2006}, and empirical work suggests that alignment patterns vary with %proficiency and 
interactional context \citep{Costa01062008, KIM2023103007}. In the present one-shot design, we therefore treat uptake as short-term mining or alignment, not as evidence of durable learning; repeated alignment may provide affordances for learning over time, but our data measure immediate reuse opportunities rather than acquisition.

Measuring alignment at scale remains challenging. However, computational dialogue research offers measures ranging from global similarity to targeted accounts of repetition and construction reuse \citep{REITTER201429}. Work in human computer interaction further shows that people align their language with spoken dialogue systems, and that the strength and form of this alignment can vary with system framing and design \citep{cowan2015lexicalalignment, spillner2021linguisticalignment}. Educational natural language processing has also moved toward turn-linked measures that capture responsiveness beyond surface similarity \citep{demszky2021measuring}. Recent work similarly demonstrates the feasibility of automated, language-based measurement in educational settings, including LLM-based detection of learning-relevant constructs in verbal protocols \citep{2024.EDM-long-papers.13}. At the same time, these approaches foreground familiar concerns about construct validity and fairness in model-based inference \citep{2025.EDM.long-papers.158}. Building on these threads, we operationalize lexical and morphosyntactic uptake as a direction-sensitive form of alignment, measuring the extent to which learners reuse lexical items and morphosyntactic patterns introduced by the interlocutor in the immediately preceding context.

\paragraph{Learner affect and attitudes} CALL studies of chatbot and LLM-based speaking practice frequently include self-report measures---perceived usefulness, enjoyment, comfort, willingness to communicate---since these perceptions shape sustained engagement \citep{huang2022_chatbots_review, fryer2006bots}. Recent work shows that some conversational AI designs can increase willingness to communicate and reduce speaking anxiety \citep{Wang_2025_understanding_gen_AI_chatbots_social_anxiety}, and that perceived enjoyment and satisfaction predict intentions to continue using AI tools for speaking practice \citep{Huang_English_speaking_with_ai}. However, these attitudinal outcomes are rarely connected to fine-grained interaction traces, leaving open which moment-by-moment behaviors drive learner perceptions---a gap we address by relating survey responses to transcript-derived interaction metrics.

\section{Data Collection}
\label{sec:study}

\subsection{Task and Stimuli}

Participants completed a spot-the-difference task, a cooperative information-gap activity widely used in communicative language teaching \citep{Pica_Kang_Sauro_2006}. A human artist created a ``haunted house'' scene containing eight rooms, each drawn in two matched versions with systematic differences in object placement (e.g., repositioned ghosts, added or removed objects). Participants communicated in the L2 language to identify these differences. The eight rooms were roughly matched for complexity and divided into two sets of four; one set was used for the human dialogue condition and one for the AI dialogue condition. Figure~\ref{fig:experiment-flow} provides an overview of the study design.
%matched for elicitation complexity ($M = \text{[X]}$\,s vs.\ $M = \text{[X]}$\,s in monologue duration)
%One of the pair used image A, and the other used image B for tasks 1 and 2.

\subsection{Conditions and Procedure}

The study used a within-participant design with four speaking conditions along two dimensions: (i)~task format (monologue vs.\ dialogue) and (ii)~partner type (human peer vs.\ AI model). %As shown in Figure~\ref{fig:design}, each participant completed two blocks: within each block, a monologue preceded a dialogue on the same room set. Human dialogue always used room set A/B and AI dialogue always used C/D; the block order was counterbalanced across participants.

Participants were assigned to a counterbalanced design, with one group completing the human–human condition first and the other completing the human–AI condition first. In the human-human condition, participants first completed a monologue describing either image A or image B, and then engaged in a dialogue about the same room set with a human partner who had previously described the complementary version. In the human-AI condition, all participants first described image C, followed by a dialogue with the AI system, which was provided a textual description of image D.

In human dialogues, participants were paired with a classmate and completed the task {face-to-face or on Zoom}; no constraints on turn-taking or strategy were imposed. AI dialogues were conducted through a web application with audio input. This modality asymmetry is representative of how each mode would typically be deployed, but co-presence in the human condition provides non-verbal coordination cues unavailable in the AI condition. Each session lasted approximately 90 minutes including a background survey, vocabulary assessment, the four speaking tasks, and a post-task survey. Data were collected between October and December 2025.

All procedures were approved by the institutional review boards at the participating sites. Before completing the background survey and speaking tasks, participants provided informed consent for audio recording and transcript-based analysis; recordings without valid consent were excluded.

\subsection{AI Interlocutor System}
%Because direct image input produced excessive hallucination during piloting, we provided the AI with manually verified English-language text descriptions of each room, written by the research team and checked against the images for accuracy. %At session initialization, all room descriptions were included in the system prompt\footnote{An anonymized repository containing the full LLM prompt is available at \url{https://osf.io/9kxsu/overview?view_only=ff7b5f764f1647a788c8541bc0b3fc31}.}; at each room transition, the current room's description was re-sent. The AI retained full conversation history within a session.

We developed the AI system using OpenAI's \texttt{gpt-realtime} model\footnote{Specifically, \texttt{gpt-realtime-2025-08-28}}, selected for its native speech-to-speech capability (eliminating separate ASR/TTS latency), support for a plurality of languages, and real-time turn-taking via server-side voice activity detection. Because direct image input produced excessive hallucination during piloting, we provided the AI with manually verified English-language text descriptions of each room, written by the research team and checked against the images for accuracy. This choice kept the AI grounded in the task materials but creates a stimulus-representation asymmetry: human partners viewed images directly, whereas the AI received textual descriptions of the rooms.

At session initialization, all room descriptions were included in the system prompt\footnote{An anonymized repository containing the full LLM prompt is available at \url{https://osf.io/9kxsu/overview?view_only=ff7b5f764f1647a788c8541bc0b3fc31}. }; at each room transition, the current room's description was re-sent. The AI retained full conversation history within a session.

The system parameters were set as follows: voice \texttt{alloy}, temperature $0.8$, max response tokens $150$. Turn-taking used server-side VAD with an open microphone (prefix padding 300\,ms, silence threshold 200\,ms). After pilot testing, we found it useful to instruct the AI model to use informal register, take short turns (one to two sentences), allow time for learner responses, and avoid excessive praise. Despite these constraints, the AI consistently occupied a larger share of the conversational floor than human partners.

\subsection{Participants}

All participants were enrolled in second- or third-year university German courses at one of four sites: two Flemish universities in Belgium ($n = 24, 34$), a Midwestern US university ($n = 14$), and a Pacific Northwest US university ($n = 14$), totaling $N = 86$. After excluding non-consented, failed, and incomplete recordings, $N = 78$ participants remained, yielding
78 AI dialogues and 39 human-human conversations (one
per dyad). Per-analysis sample sizes vary by minimum-turn thresholds and are reported with each analysis.
Approximately 50\% of participants listed Dutch as their sole L1, English was the second largest group (34.5\%), and the remaining 15\% comprised speakers of Romanian, Vietnamese, Spanish, Albanian, Hindi, and Ndebele, several of whom reported bilingual combinations with Dutch.

\subsection{Transcription}
We analyze diarized transcripts rather than raw audio, a deliberate choice favoring scalability: transcript-based measures avoid prosodic confounds across L1 backgrounds and recording environments, require no manual coding, and yield an easily replicable pipeline.

All sessions were transcribed using ElevenLabs \texttt{scribe\_v2} with automatic speaker diarization (language: \texttt{deu}), producing word-level timestamps and speaker attribution. German-speaking authors reviewed all transcripts, verifying speaker labels, trimming non-task speech at recording boundaries, and confirming coherence with assigned room content. Word-level ASR errors were not systematically corrected. In three AI-dialogue transcripts, mid-stream speaker label swaps were detected via an automated disfluency-rate classifier (the AI produces zero filled pauses) and corrected manually.

\section{Methods}\label{sec:methods}

From the diarized, timestamped transcripts we extract two complementary sets of metrics. \emph{Interactional fluency} (\S\ref{sec:temporal-dynamics}) captures how speakers coordinate in real time -- how long they hold the floor, how quickly they respond, and how much they hesitate within turns. \emph{Linguistic uptake} (\S\ref{sec:linguistic-uptake}) captures what speakers do with the language they hear -- whether they adopt vocabulary or reproduce morphosyntactic patterns from the interlocutor's speech. The first captures \emph{how} learners interact; the second captures \emph{what language} they take away from the interaction.

\subsection{Interactional Fluency}
\label{sec:temporal-dynamics}

Dialogue involves continuous coordination over who speaks, for how long, and when transitions occur. These properties can differ systematically between human and AI partners, shaping the kind of practice an interaction affords. We capture this through seven features extracted from word-level timestamps, organized into two groups.% (Table~\ref{tab:fluency-results}).

\paragraph{Conversation structure} Four features describe the overall shape of a dialogue. \emph{Conversation length}: total session duration in seconds. \emph{Turn count}: number of participant turns. \emph{Mean turn duration}:  average length of a participant turn in seconds. \emph{Words per turn}: average words per participant turn. Together, these indicate whether the dialogue consists of many short exchanges or fewer, longer stretches of speech.

\begin{figure*}[!t]
\centering
\includegraphics[width=\textwidth]{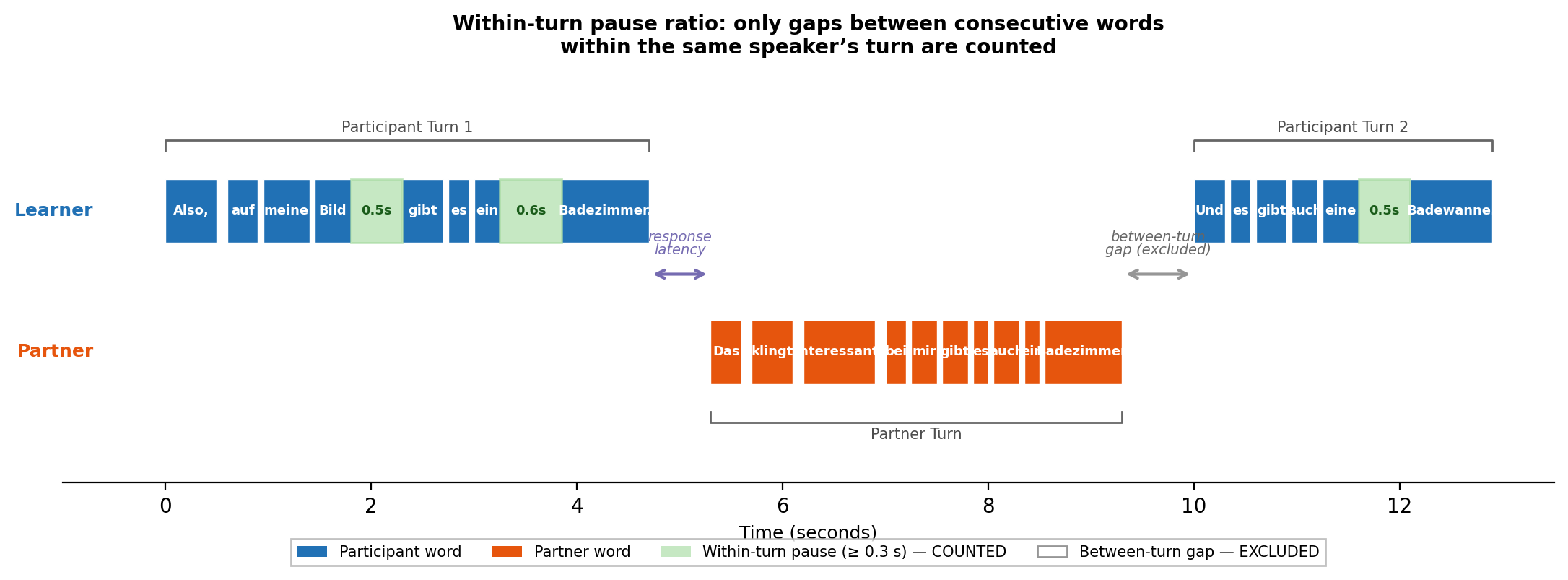}
\caption{Illustration of temporal metrics extracted from word-level timestamps. Blue blocks are participant words; orange blocks are partner words. \emph{Within-turn pause ratio} counts only gaps $\geq 0.3$\,s between consecutive words within the same speaker's turn (green). Gaps between speakers' turns (gray) are excluded from the pause ratio but are used to compute \emph{response latency} (purple). Turn boundaries are reconstructed by grouping consecutive words from the same speaker.}
\Description{Schematic of temporal metrics from word-level timestamps. Participant words and partner words are shown as separate blocks, with within-turn pauses, between-speaker gaps, response latency, and turn boundaries labeled.}
\label{fig:temporal_pause}
\end{figure*}

\paragraph{Turn-taking dynamics}
Three features capture how speakers manage transitions and share the floor. \emph{Response latency}: the gap between the partner's
last word and the participant's first word; only gaps
within $[0, 10]$\,s are retained. \emph{Floor ratio}: the proportion of session time the participant is speaking. \emph{Within-turn pause ratio}: the proportion of a participant's turn spent in silence ($\geq 0.3$\,s gaps between words), reflecting online planning difficulty. Figure~\ref{fig:temporal_pause} illustrates how within-turn pauses are distinguished from between-turn gaps.

Session-level metrics (conversation length, turn count) were compared using paired $t$-tests ($n = 78$). Per-turn metrics use sessions with $\geq 8$ participant turns, yielding unbalanced samples; these were analyzed using linear mixed-effects models with condition as a fixed effect and participant as a random intercept.

% Session-level metrics (conversation length, turn count) use the full sample ($n = 78$ per condition). Per-turn metrics require $\geq 8$ participant turns ($n_{\text{AI}} = 63$, $n_{\text{Human}} = 72$).

\subsection{Linguistic Uptake}
\label{sec:linguistic-uptake}

\begin{figure*}[htbp]
    \centering
    \includegraphics[width=.75\textwidth]{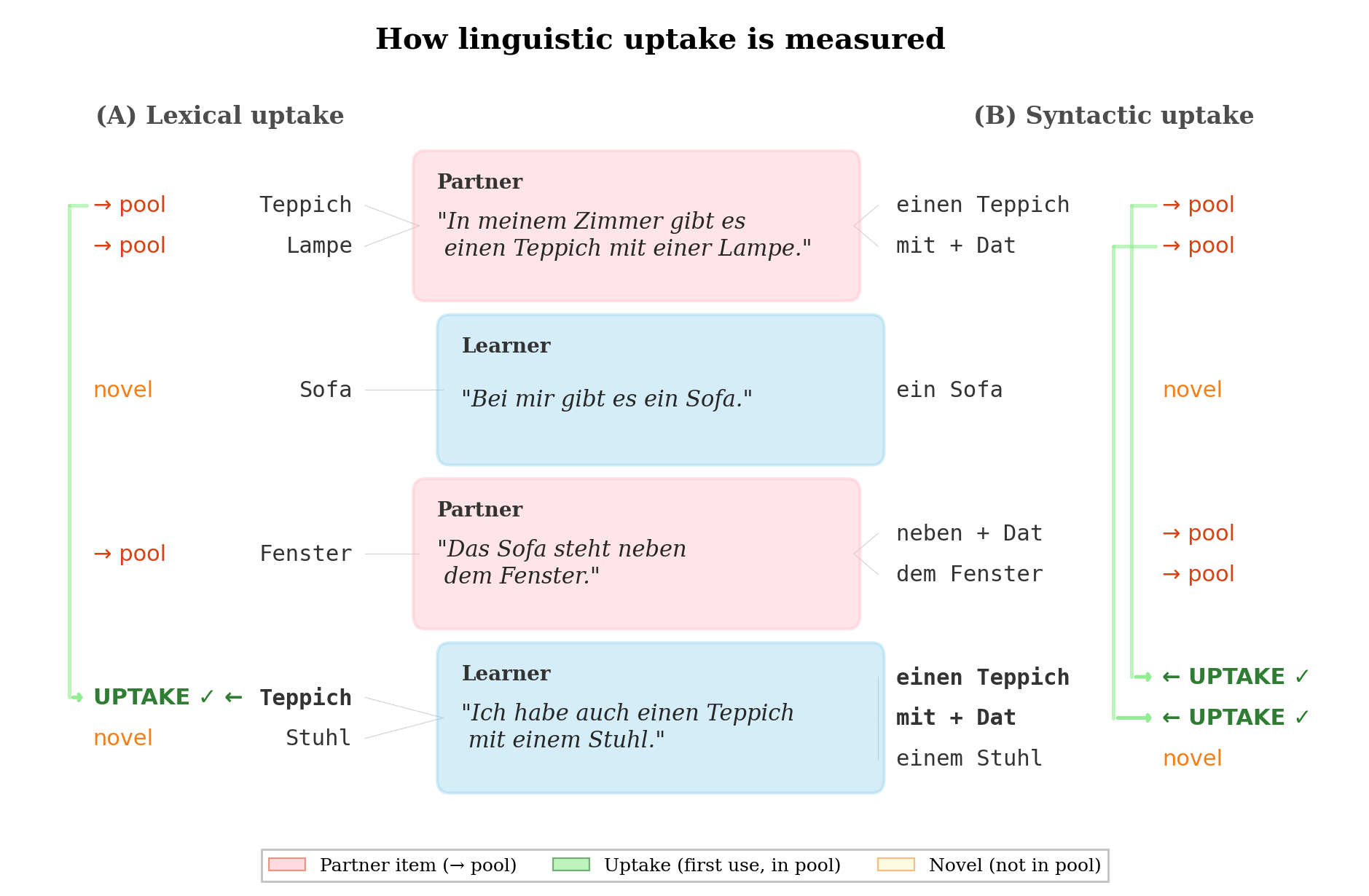}
    \caption{Illustration of linguistic uptake measurement at two
    levels, applied to the same example dialogue
    ($N_{\text{AI}} = 78$, $N_{\text{Human}} = 78$ in the full
    dataset). \textbf{(A)}~Lexical uptake: content words used by the
    partner accumulate in a pool; when the learner first produces a
    matching word, it is counted as uptake. \textbf{(B)}~Syntactic
    uptake: morphosyntactic templates---determiner--noun agreement
    (e.g., \emph{einen Teppich}, masculine~accusative) and
    preposition case government (e.g., \emph{mit} +
    dative)---are tracked analogously. Green arrows trace each uptake
    event to its source in the partner's speech. Only the learner's
    first use of each item counts (first-use constraint); repeated
    uses are not double-counted. For syntactic templates, grammatical
    correctness is additionally validated against a noun-gender
    lexicon (e.g., \emph{Teppich} $\to$ masculine).}
\Description{Schematic of linguistic uptake measurement. Partner vocabulary and morphosyntactic templates accumulate as available input, and the learner's first matching use is counted as lexical or syntactic uptake.}
    \label{fig:uptake}
\end{figure*}

Interactive alignment theory \citep{pickering2004interactive} predicts
that interlocutors converge on shared linguistic representations
during dialogue, from lexical choices to syntactic structures. We
operationalize this as {linguistic uptake}: whether L2 learners
adopt linguistic material from their interlocutor's speech. 
Figure~\ref{fig:uptake} presents an illustration of how uptake is estimated. We measure uptake at two complementary
levels:

\noindent \textbf{Lexical uptake} tracks whether learners adopt vocabulary items (lemmas) introduced by the partner, capturing
      breadth of vocabulary activation across the full session.\\
    \noindent \textbf{Syntactic uptake} tracks whether learners reproduce morphosyntactic patterns (e.g., determiner--noun agreement, prepositional case government) from the partner's prior speech, capturing structural alignment within a dialogue.

Both measures impose a \textbf{first-use constraint}, such that each linguistic unit is scored only when it first appears in the learner’s speech. This operationalization follows an input-mining paradigm in L2 research, where learners' later production is interpreted relative to prior independent production and intervening input \citep{Hoang_Boers_2016}. In our design, the monologue preceding each dialogue provides a learner-specific baseline for lexical uptake, while syntactic uptake applies the same conservative first-use logic within dialogue by crediting templates only after prior partner use within a bounded recency window. We use uptake/mining to mean short-term reuse of available input, not acquisition; repeated task performance and the shared visual context may also contribute to some cases.

\subsubsection{Lexical Uptake}

The lexical uptake algorithm uses a \textbf{global timeline}
approach, tracking each lemma across all four tasks in sequential
order. The counterbalanced design creates two orderings:

\begin{itemize}
    \item \textbf{Group~1} ($n = 46$): Monologue~A/B $\rightarrow$
      Human dialogue $\rightarrow$ Monologue~C $\rightarrow$
      AI dialogue
    \item \textbf{Group~2} ($n = 32$): Monologue~C $\rightarrow$
      AI dialogue $\rightarrow$ Monologue~A/B $\rightarrow$
      Human dialogue
\end{itemize}

The participant's first monologue establishes a \textbf{baseline
vocabulary}. Words are tokenized, filtered (removing determiners, 49
pre-task vocabulary items, 153 English code-switching tokens, and
words $\leq 2$ characters), and lemmatized using spaCy's
\texttt{de\_core\_news\_sm} model. Words appearing on the
Goethe-Institut A1 wordlist
\citep{goethe-institute-A1-vocab-list-perlmann-balme2012},
approximately 714 high-frequency items, are also excluded as
uninformative adoption candidates. A lemma counts as
\textbf{adopted} if: (1)~the interlocutor produced it before the
participant, (2)~it is not in the participant's baseline, and (3)~it
has not already been attributed to another source
(\textbf{first-source-wins} rule). We use ``adopted'' rather than
``learned'' because we are measuring activation through interactive
exposure, not acquisition.

\begin{table*}[!t]
\centering
\caption{Target morphosyntactic patterns for syntactic uptake.}
\label{tab:target-patterns}
\begin{tabular}{lll}
\hline
\textbf{Pattern} & \textbf{Example} & \textbf{What it captures} \\
\hline
\textsc{det-noun} & \emph{der Tisch} (Masc/Nom/Sg) &
  Det--noun gender/case/number agreement \\
\textsc{adp-case} & \emph{mit dem Hund} (mit $\rightarrow$ Dat) &
  Preposition case government \\
\textsc{adj-noun} & \emph{gro\ss es Gem\"alde} (Neut/Nom/Sg) &
  Attributive adjective agreement \\
\textsc{contraction} & \emph{im} (= \emph{in} + \emph{dem}, Dat) &
  Preposition--article fusion \\
\hline
\end{tabular}
\end{table*}

Adoption is measured cumulatively without a recency window, consistent
with accounts of lexical entrainment as a \emph{historical}
phenomenon in which interlocutors maintain partner-specific conceptual
precedents beyond immediate adjacency
\citep{brennan1996conceptual}, and with L2 alignment work that omits
distance constraints under an interaction-as-learning framing
\citep{MICHEL2022102930}. Because the AI interlocutor produces
substantially more speech than human partners, we report both raw
adoption counts and adoption rates controlled for verbosity by
dividing by the number of unique novel lemmas each interlocutor
introduced beyond the participant's baseline (\textbf{pool size}). To
test whether interaction type predicts adoption beyond input volume,
we fit a mixed-effects regression with condition, pool size, and
learner word count as fixed effects and participant as a random
intercept.

\subsubsection{Syntactic Uptake}

Syntactic uptake measures whether learners reproduce morphosyntactic patterns from the interlocutor's speech \textbf{within a single dialogue}. We extract four pattern types that constitute the core of German nominal morphology (see Table~\ref{tab:target-patterns}): determiner--noun agreement, prepositional case government, attributive adjective agreement, and preposition--article contractions. These are pervasive in the data ($M = 86$ learner-produced template instances per AI dialogue, $113$ per human dialogue) and target well-documented areas of L2 German difficulty.

\paragraph{Extraction pipeline} Each turn is lightly preprocessed (removing filled pauses such as \emph{\"ah} and \emph{\"ahm}) and parsed with the Stanza German UD pipeline \citep{qi-etal-2020-stanza}. Each extracted template has an \textbf{anchored} representation that includes the noun lemma, surface form, and morphological features (e.g., \textsc{det\_}\allowbreak\textsc{noun}|\allowbreak\emph{tisch}|\allowbreak\emph{dem}|\allowbreak Masc|\allowbreak Dat|\allowbreak Sg), tying uptake to specific lexical items rather than abstract morphological categories.

\paragraph{Correctness checking} For \textsc{det-noun} and \textsc{adj-noun} templates, we verify gender agreement against a lexicon of 91,270 German nouns. For \textsc{adp-case}, correctness is checked for strict-case prepositions (e.g., \emph{mit} $\rightarrow$ dative); two-way prepositions (e.g., \emph{auf}, \emph{in}), which can take dative or accusative case, are included in uptake matching but excluded from correctness analysis, as case selection depends on semantic context that cannot be determined from syntax alone. Contractions are inherently correct.

\begin{figure*}[!t]
\centering
\includegraphics[width=\textwidth]{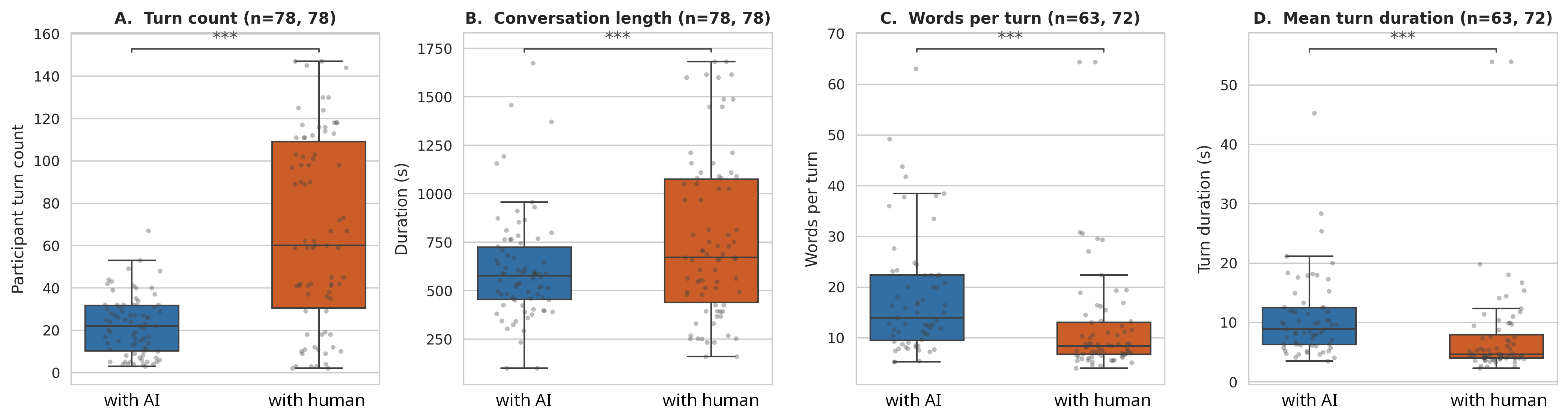}
\caption{Conversation structure: AI vs.\ Human partner. (A)~Participant turn count and (B)~conversation length use all sessions ($n =78$ per condition). (C)~Words per turn and (D)~mean turn duration use sessions with $\geq 8$ participant turns ($n_{\text{AI}} =63$, $n_{\text{Human}} = 72$). Box plots show median and interquartile range; individual sessions are overlaid as points. $^{***}p < .001$.}
\Description{Box plots comparing conversation structure with AI and human partners. Human dialogues have more participant turns and longer total conversation time, while AI dialogues have more words per turn and longer mean participant turns.}
\label{fig:structure}
\end{figure*}

\begin{figure*}[!t]
\centering
\includegraphics[width=.85\textwidth]{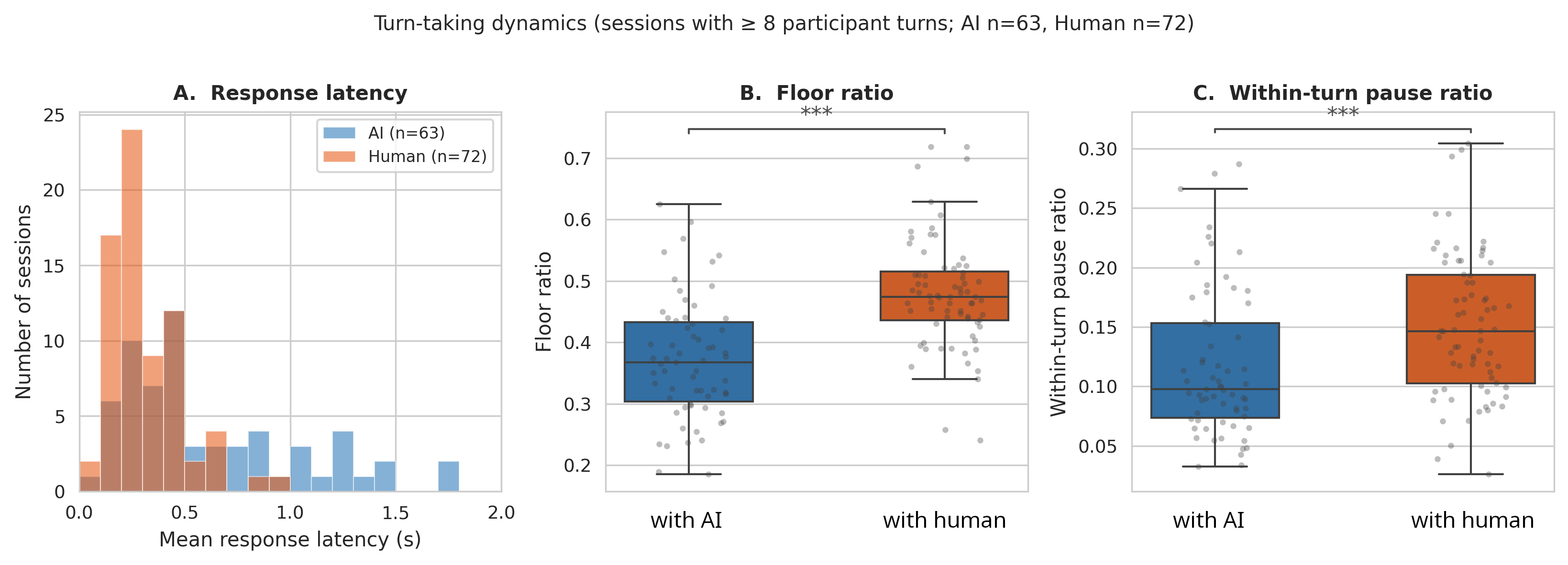}
\caption{Turn-taking dynamics: AI vs.\ Human partner (sessions with $\geq 8$ participant turns). (A)~Overlapping histograms of mean response latency show clear separation between conditions. (B)~Floor ratio: learners hold the floor for a smaller share of session time with the AI. The dashed line marks equal floor time ($0.5$). (C)~Within-turn pause ratio: learners pause \emph{less} within turns when speaking with the AI. $^{***}p < .001$.}
\Description{Plots comparing turn-taking dynamics with AI and human partners. Learners show longer response latency, lower floor ratio, and lower within-turn pause ratio in AI dialogue than in human dialogue.}
\label{fig:turn-taking}
\end{figure*}

\paragraph{Uptake scoring} On the participant's first production of
a template, we check whether it appeared in the source's prior
speech. Unlike lexical adoption, syntactic uptake uses a recency
window, reflecting the lag-sensitivity of structural priming
documented in both L1 \citep{HARTSUIKER2008214} and L2 task-based
interaction
\citep{jbp:/content/journals/10.1075/itl.17021.dao}, where
prime--target contingency is typically coded within a bounded span
\citep{MICHEL2022102930}. Because AI turns are longer and
syntactically denser than human turns, a fixed turn-count window
captures unequal amounts of input across conditions. We therefore
compare three window operationalizations: \textbf{turn-based} (last
3~source turns), \textbf{time-based} (last 15 seconds of source
speech), and \textbf{template-count} (last 15 source templates,
explicitly equalizing input volume). All results are tested via
mixed-effects regression with condition, source pool size, and
learner unique template count as fixed effects and participant as a
random intercept.

% \begin{table*}[!t]
% \centering
% \caption{Interactional dynamics: AI vs.\ Human partner. All comparisons are statistically significant (Welch's $t$-tests; $d$ = Cohen's $d$; {*}\,$p < .05$, {**}\,$p < .01$, {***}\,$p < .001$). Per-turn metrics ($^{\ddagger}$) use sessions with $\geq 8$ participant turns; session-level metrics use all sessions.}
% \label{tab:fluency-results}
% \begin{tabular}{llccccccl}
% \hline
% Domain & Metric & AI $M$ ($SD$) & $n$ & Human $M$ ($SD$) & $n$ & $t$ & $p$ & $d$ \\
% \hline
% \multirow{4}{*}{Structure}
%  & Conversation length (s)          & 608 (271)    & 78 & 764 (417)    & 78 & $-2.78$ & .006**           & $-0.44$ \\
%  & Turn count                       & 22.2 (13.9)  & 78 & 66.5 (44.2)  & 78 & $-8.43$ & ${<}.001$***     & $-1.35$ \\
%  & Mean turn duration (s)$^{\ddagger}$  & 10.8 (6.9)   & 63 & 7.7 (8.8)    & 72 & 2.29    & .024*            & 0.39 \\
%  & Words per turn$^{\ddagger}$          & 18.0 (11.9)  & 63 & 12.1 (11.0)  & 72 & 2.97    & .004**           & 0.51 \\
%  \cline{2-9}
% \addlinespace
% \multirow{3}{*}{Turn-taking}
%  & Mean response latency (s)$^{\ddagger}$ & 0.61 (0.42)  & 63 & 0.32 (0.18)  & 72 & 5.11    & ${<}.001$*** & 0.90 \\
%  & Floor ratio$^{\ddagger}$               & 0.37 (0.10)  & 63 & 0.48 (0.09)  & 72 & $-6.61$ & ${<}.001$*** & $-1.14$ \\
%  & Within-turn pause ratio$^{\ddagger}$   & 0.12 (0.06)  & 63 & 0.15 (0.06)  & 72 & $-3.08$ & .003**       & $-0.53$ \\
% \hline
% \end{tabular}
% \end{table*}

The algorithm also detects \textbf{error--correction sequences}: the participant produces an incorrect form, the source subsequently models the correct form, and the participant later produces the corrected version.

%%%%%%%%%%% RESULTS 
\section{Results}\label{sec:results}
We present results in three parts. Interactional dynamics (\S\ref{sec:temporal-dynamics}) characterize how dialogue structure and turn-taking differ between AI and human partners. Lexical and syntactic uptake (\S\ref{sec:linguistic-uptake}) examine whether and how learners adopt language from each interlocutor type. Survey results (\S\ref{sec:survey-results}) connect interaction behavior to learner attitudes and satisfaction.

\subsection{Interactional Dynamics}

%Table~\ref{tab:fluency-results} presents the full comparison. 
We organize the findings into three domains: conversation structure, turn-taking dynamics, as well as consistency of results across four sites.

\begin{figure*}[!t]
\centering
\includegraphics[width=\textwidth]{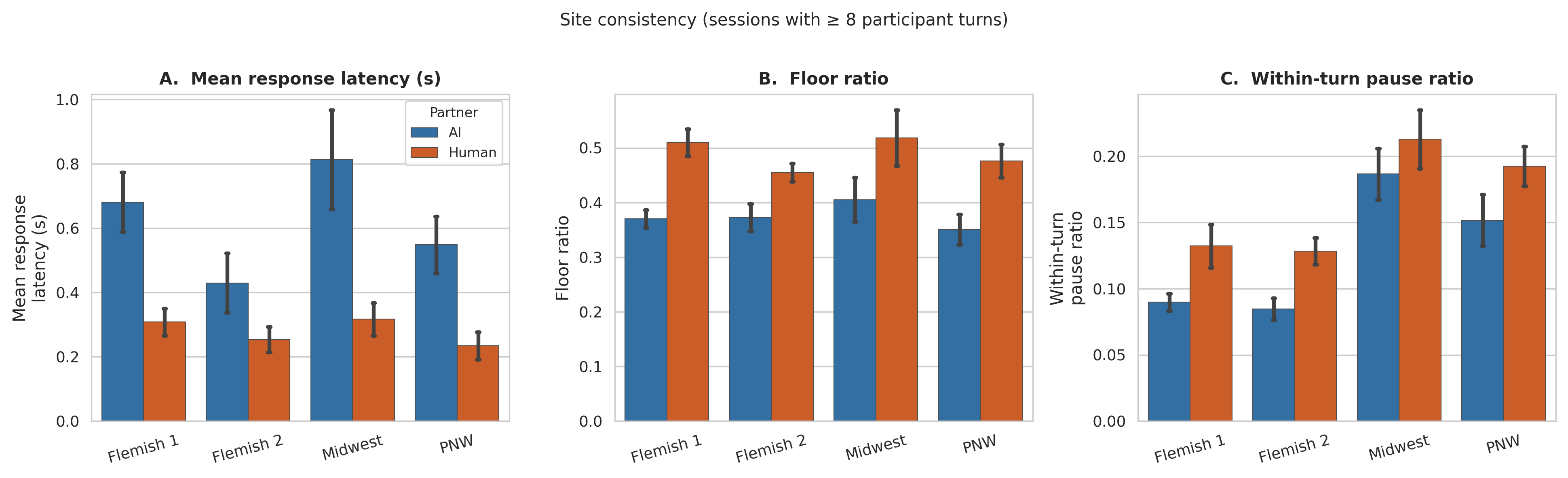}
\caption{Site consistency: the three turn-taking metrics broken down by site and partner type (sessions with $\geq 8$ turns). Error bars show $\pm 1$ SE. At every site, response latency is higher with the AI, floor ratio is lower, and within-turn pause ratio is lower. No site reverses the direction of any effect.}
\Description{Site-level comparison of turn-taking metrics by partner type. Across all four sites, response latency is higher with AI, floor ratio is lower with AI, and within-turn pause ratio is lower with AI.}
\label{fig:site-comparison}
\end{figure*}

\subsubsection{Conversation Structure}

Human-human dialogues were significantly longer ($M = 764$\,s vs.\ $608$\,s, $d = -0.44$, $p < .001$) and contained nearly three times as many turns ($M = 66.5$ vs.\ $22.2$, $d = -1.35$, $p < .001$). Human partners engaged in rapid back-and-forth exchange, while AI dialogues were characterized by fewer, longer turns. With the AI, participants produced significantly longer turns ($M = 10.8$\,s vs.\ $7.7$\,s, $d = 0.39$, $p < .001$) containing more words per turn ($M = 18.0$ vs.\ $12.1$, $d = 0.51$, $p < .001$). Learners pack more content into each turn with the AI, but the threefold reduction in turn count means that total learner output is substantially lower than in human dialogue.

Figure~\ref{fig:structure} illustrates this trade-off: human dialogues produce many more turns (A), but learners produce more words per turn with the AI (C).

\subsubsection{Turn-Taking Dynamics}

Three interrelated measures reveal qualitatively different turn-taking patterns across conditions. Figure~\ref{fig:turn-taking} visualizes the three turn-taking metrics, showing clear separation between conditions for response latency (A) and consistent differences in floor ratio (B) and within-turn pause ratio (C).

Participants took roughly twice as long to begin speaking after the AI ($M = 0.61$~s) as after a human partner ($M = 0.32$~s; $d = 0.90$, $p < .001$), indicating substantially longer response latencies in AI dialogue. Most human-partner latencies cluster below 0.3~s, while AI latencies spread between 0.2--1.0~s, likely reflecting the AI's longer, more information-dense turns.

Learners held the floor for a smaller proportion of session time when interacting with the AI ($M = 0.37$) than with a human partner ($M = 0.48$; $d = -1.14$, $p < .001$), representing one of the largest effects observed in the dataset. The AI’s longer turns account for much of this difference, as they occupy a greater share of the conversational time.

Learners paused \emph{less} within their own turns when speaking with the AI ($M = 0.12$ vs.\ $0.15$; $d = -0.53$, $p < .001$), indicating greater within-turn fluency. When learners do hold the floor with the AI, they tend to speak more fluently within each turn.

Taken together, these measures show a consistent shift in interactional organization: AI dialogue features slower turn transitions and reduced learner floor time, but more fluent production when learners hold the floor.
%The AI's patient, non-competitive interaction style appears to reduce within-turn planning pressure, allowing more continuous speech production.

\subsubsection{Site Consistency}

Figure~\ref{fig:site-comparison} shows the contrasts were consistent in direction across all four sites. Response latency was higher with the AI at every site; floor ratio was lower; within-turn pause ratio was lower. The magnitudes varied---Midwest showed the   largest response-latency difference, while Flemish~2 showed the smallest floor-ratio gap---but no site reversed the direction of any effect.

\subsection{Linguistic Uptake}

\subsubsection{Lexical Uptake}
The AI produced over twice as many novel lemmas as human partners (pool size: $M = 79.6$ vs.\ $36.4$), yielding a higher descriptive adoption rate for human partners ($16.2\%$ vs.\ $9.3\%$ of pool adopted). However, a mixed-effects regression controlling for pool size and learner word count found no significant condition effect ($\beta = -1.77$, $SE = 0.95$, $z = -1.87$, $p = .062$). Pool size ($\beta = 0.10$, $p < .001$) and learner word count ($\beta = 0.005$, $p < .001$) were the significant predictors: after accounting for input volume and learner output, condition does not reliably predict adoption.

%\paragraph{Position effects}
To disentangle partner type from task order, we compared adoption at the same sequential position. Table~\ref{tab:lexical-adoption} presents the results of this analysis. In position~2---where the depleted vocabulary pool provides a more sensitive test---AI participants adopted significantly more words ($M = 4.6$ vs.\ $2.6$; $d = 0.72$, $p = .003$). In position~1, the direction is consistent but nonsignificant ($d = 0.42$, $p = .079$).

\begin{table*}[t]
\centering
\caption{Lexical adoption by sequential position and partner type. Group~1 ($n = 46$) had the human partner first; Group~2 ($n = 32$) had the AI first. Position-matched comparisons use between-subjects Welch's $t$-tests.}
\label{tab:lexical-adoption}
\begin{tabular}{llcccccc}
\hline
\textbf{Position} & \textbf{Partner type} & \textbf{$n$} & \textbf{$M$} & \textbf{$SD$} & \textbf{$t$} & \textbf{$p$} & \textbf{$d$} \\
\hline
\multirow{2}{*}{First dialogue}  & Human (Group~1) & 46 & 8.6 & 5.5  & \multirow{2}{*}{1.79} & \multirow{2}{*}{.079} & \multirow{2}{*}{0.42} \\
                                 & AI (Group~2)    & 32 & 11.2 & 6.8 &                       &                       &                       \\
\hline
\multirow{2}{*}{Second dialogue} & Human (Group~2) & 32 & 2.6  & 2.8  & \multirow{2}{*}{3.12} & \multirow{2}{*}{.003**} & \multirow{2}{*}{0.72} \\
                                 & AI (Group~1)    & 46 & 4.6  & 2.8  &                       &                         &                       \\
\hline
\end{tabular}
\end{table*}

% \paragraph{Summary.} AI dialogue exposes learners to roughly twice the novel vocabulary, and learners adopt more words in absolute terms. However, this advantage is attributable to the AI's greater verbosity: after controlling for pool size and learner output, condition does not reliably predict adoption ($p = .062$). The AI creates more lexical \emph{exposure}; it does not appear to create qualitatively different conditions for lexical \emph{uptake}.

\subsubsection{Syntactic Uptake}
A total of 156 participant-level dialogue records were analyzed: 78~AI dialogue records, and 78~human-partner records derived from 39 human-human conversations. The results are summarized in Table~\ref{tab:syntactic-uptake}. Learners' cumulative uptake rate was higher with the AI ($M = .16$, $SD = .11$) than with human partners ($M = .08$, $SD = .06$; $d = 0.93$), but as with lexical adoption, the AI also provided a substantially larger source pool ($M = 155$ vs.\ $90$ unique templates; $d = 1.35$). After controlling for pool size and learner unique template count, cumulative uptake showed a positive but nonsignificant trend ($\beta = 1.55$, $SE = 0.92$, $z = 1.69$, $p = .090$). Recency-windowed uptake told a different story: the AI advantage was significant across all three operationalizations---turn-based ($\beta = 3.18$, $p < .001$), time-based ($\beta = 3.04$, $p < .001$), and template-count ($\beta = 2.84$, $p = .001$)---including the window that explicitly equalizes input volume. This confirms that the AI's short-term priming advantage reflects genuine structural priming, not merely a larger input pool.

\begin{table*}[t]
\centering
\caption{Syntactic uptake by condition. Descriptive statistics (Welch's $t$-tests) and mixed-effects regression results ($\beta$ = condition coefficient, AI${}= 1$; models control for source pool size and learner unique templates with participant random intercept).}
\label{tab:syntactic-uptake}
\begin{tabular}{lcccccccc}
\hline
\textbf{Measure} & \textbf{AI $M$ ($SD$)} & \textbf{Human $M$ ($SD$)} & \textbf{$t$} & \textbf{$d$} & \textbf{$\beta$} & \textbf{$SE$} & \textbf{$z$} & \textbf{$p$} \\
\hline
Cumulative uptake (rate)      & .16 (.11)  & .08 (.06) & 5.84  & 0.93 & --- & --- & --- & ---\\
Cumulative uptake (count)     & 7.3 (5.6)  & 5.4 (4.6) & 2.28  & 0.36 & 1.55 & 0.92 & 1.69 & .090\\
Recency: 3 turns (count)      & 6.3 (5.2)  & 2.9 (2.9) & 4.98  & 0.80 & 3.18 & 0.86 & 3.69 & ${<}.001$*** \\
Recency: 15\,s (count)        & 5.7 (5.0)  & 2.7 (2.7) & 4.65  & 0.74 & 3.04 & 0.84 & 3.61 & ${<}.001$*** \\
Recency: 15 templates (count) & 5.2 (5.3)  & 3.2 (2.8) & 3.00  & 0.48 & 2.84 & 0.87 & 3.27 & .001** \\
\hline
\end{tabular}
\end{table*}

Correctness rates were comparable: AI $M = .82$ ($SD = .13$) vs.\ Human $M = .80$ ($SD = .13$); $p = .316$, $d = 0.16$. Higher uptake with the AI does not come at the cost of morphological accuracy. %\paragraph{Error--correction sequences} 
Correction sequences were numerically more frequent with human partners (total: 38 vs.\ 27), but per-participant rates did not differ significantly ($M = 0.49$ vs.\ $0.35$; $p = .215$). 

\subsection{Survey Analysis}
\label{sec:survey-results}
Of the 78 participants, 55 completed matched pre- and post-surveys; 52 had complete data for the directional sentiment composite used in the paired $t$-test below. These data allow us to assess how learner attitudes toward AI shift after direct experience and whether behavioral features of the interaction predict satisfaction.

\textbf{Baseline attitudes.} 
Participants entered with predominantly skeptical attitudes toward AI: 60\% described themselves as skeptical or unsure, and only 14.5\% as enthusiastic adopters. AI-specific Likert items scored near or below neutral (perceived effectiveness $M = 3.22$, enjoyment $M = 2.73$), while general speaking practice was highly valued ($M = 4.56$). Notably, relaxation with AI ($M = 3.25$) already exceeded relaxation with proficient speakers ($M = 2.82$), and the lowest-scoring item, having enough opportunities to speak with proficient speakers ($M = 2.67$), points to a perceived access gap (see Table~\ref{tab:baseline_key}).

\begin{table}[!t]
\centering
\caption{Key baseline attitudes: AI vs. Human peers (N=55). All items rated on a 5-point Likert scale (1 = Strongly Disagree, 5 = Strongly Agree).}
\label{tab:baseline_key}
\begin{tabular}{lcc}
\hline
\textbf{Item} & \textbf{$M$} & \textbf{$SD$} \\
\hline
Practicing speaking is valuable & 4.56 & 0.69 \\
AI effectiveness & 3.22 & 0.76 \\
AI enjoyment & 2.73 & 0.89 \\
Relaxation: AI & 3.25 & 1.02 \\
Relaxation: Proficient speakers & 2.82 & 1.19 \\
Access to proficient speakers & 2.67 & 1.11 \\
\hline
\end{tabular}
\end{table}

\textbf{Post-experience shift.}
Attitudes improved significantly after the study: a paired $t$-test on directional sentiment scores showed a mean increase of $+0.48$ ($t(51) = -4.58$, $p < .0001$, $d = 0.64$), with 75\% of participants becoming more positive. Post-study, 67.3\% liked the AI interaction, 81.8\% agreed AI can be useful for L2 speaking practice, and 34.5\% reported learning new vocabulary.

\begin{table*}[!t]
  \centering
  \small
\setlength{\tabcolsep}{4pt}
  \caption{OLS regressions predicting post-interaction satisfaction and attitude change from behavioral
  features of the AI dialogue ($n = 55$). Site is dummy-coded with Flemish~1 as the reference category.
  {*}\,$p < .05$, {**}\,$p < .01$.}
  \label{tab:survey-regression}
  \begin{tabular}{l ccc ccc ccc}
  \hline
   & \multicolumn{3}{c}{Satisfaction}
 & \multicolumn{3}{c}{Attitude change}
 & \multicolumn{3}{c}{Satisfaction (uptake)} \\
\cmidrule(lr){2-4} \cmidrule(lr){5-7} \cmidrule(lr){8-10}
%  & \multicolumn{3}{c}{(behavioral predictors)}
%  & \multicolumn{3}{c}{(behavioral predictors)}
%  & \multicolumn{3}{c}{(uptake predictors)} \\
% \cmidrule(lr){2-4} \cmidrule(lr){5-7} \cmidrule(lr){8-10}

  \textbf{Predictor} & $\beta$ & $SE$ & $p$ & $\beta$ & $SE$ & $p$ & $\beta$ & $SE$ & $p$ \\
  \hline
  Intercept                  & 4.01  & 0.54 & ${<}.001$** & 2.10  & 0.46 & ${<}.001$** & 3.34  & 0.51 &
  ${<}.001$** \\
  Pre-AI attitude            & 0.36  & 0.13 & .006**      &       &      &             & 0.31  & 0.14 & .030*
        \\
  Within-turn pause ratio    & $-4.78$ & 1.71 & .008**    & $-5.91$ & 2.07 & .006**    &       &      &
        \\
  Floor ratio                & $-0.95$ & 0.92 & .307      & $-1.16$ & 1.12 & .306      &       &      &
        \\
  Lexical adoption rate      &       &      &             &       &      &             & 1.22  & 1.62 & .454
        \\
  Syntactic uptake rate      &       &      &             &       &      &             & $-0.39$ & 1.08 & .717
        \\
  Site: Flemish 2            & $-0.53$ & 0.26 & .046*     & $-0.22$ & 0.31 & .482      & $-0.58$ & 0.27 &
  .038*   \\
  Site: Midwest              & $-0.44$ & 0.35 & .209      & $-0.18$ & 0.42 & .667      & $-0.98$ & 0.32 &
  .004**  \\
  Site: PNW                  & $-0.34$ & 0.26 & .196      & $-0.04$ & 0.31 & .907      & $-0.60$ & 0.28 &
  .037*   \\
  \addlinespace
  \hline
  $R^2$ (Adj.\ $R^2$)       & \multicolumn{3}{c}{.43 (.36)} & \multicolumn{3}{c}{.24 (.16)} &
  \multicolumn{3}{c}{.33 (.25)} \\
  $F$                        & \multicolumn{3}{c}{6.03**}    & \multicolumn{3}{c}{3.10*}     &
  \multicolumn{3}{c}{3.98**}    \\
  \hline
  \end{tabular}
  \end{table*}

\textbf{Behavioral predictors of satisfaction.}
To connect attitudinal outcomes to interaction behavior, we fit OLS regressions predicting post-satisfaction and attitude change from behavioral features of the AI dialogue. Table~\ref{tab:survey-regression} presents the results of survey regressions, fit using ordinary least squares (\texttt{statsmodels}; \cite{seabold2010}) with site as a categorical covariate.

Pre-study attitude predicted post-satisfaction ($\beta = 0.36$, $p = .006$), but the strongest behavioral predictor was within-turn pause ratio ($\beta = -4.78$, $p = .008$): learners who paused more during AI turns, reflecting greater online planning difficulty, reported lower satisfaction and smaller attitude gains ($\beta = -5.91$, $p = .006$). Floor ratio was not significant, suggesting that \emph{how fluently} learners spoke mattered more than \emph{how much} they spoke. Neither lexical adoption rate nor syntactic uptake rate predicted satisfaction ($p > .45$), indicating that the subjective experience of AI interaction is shaped by production fluency rather than by receptive alignment.

\section{Discussion}
\label{sec:discussion}
The fluency, uptake, and survey results converge on one conclusion: AI and human dialogue create distinct but complementary conditions for L2 practice.

\textbf{Monologue-in-dialogue.}
AI dialogue structurally approximates supported monologue. Learners produce sustained output with fewer, longer turns, slower responses, and less floor competition, while producing more fluent speech within each turn. The AI's patient, non-competitive style shifts learners from reactive turn management to more deliberate, content-focused production. Human dialogue, by contrast, is rapid and interactive: frequent turns, fast responses, balanced floor time, and more within-turn hesitation. 
The interactional work of peer conversation (collaborative reference-building, repair negotiation, and affiliative floor management) is largely absent from AI dialogue.
This is not a failure of the AI to ``be human'' but a description of what the interaction actually is.

\textbf{What the AI scaffolds.}
The AI's verbose, syntactically regular speech, in this specific model-prompt-task configuration, creates favorable conditions for short-term structural alignment. 
Overall, AI dialogue exposes learners to more input and elicits more uptake in absolute terms.
The deeper finding is that the AI's syntactic priming advantage survives all controls for input volume, pointing to a qualitative difference: complete, well-formed constructions without the fragments and restarts of peer speech yield cleaner targets for short-term reuse. At the lexical level, the AI provides greater exposure but not more efficient uptake per unit of input ($p = .062$ after controls). The AI's distinctive pedagogical contribution is thus specifically structural. This is particularly notable given that meaning-focused tasks like spot-the-difference typically direct learner attention toward vocabulary rather than grammar \citep{garcia2016efl}; the syntactic priming observed here operates implicitly, suggesting a potential affordance of AI interaction worth investigating in future work.

\textbf{What human partners scaffold.}
Human dialogue exercises a different set of capacities. The rapid turn exchange, floor competition, and negotiation sequences characteristic of peer interaction create conditions for developing interactional competence, skills not exercised in AI dialogue. The information-gap task produces its richest interactional outcomes with a partner who \emph{cannot} resolve gaps efficiently. Whether human dialogue also promotes more durable lexical learning through interactional grounding remains a question our data suggest but cannot confirm.

\textbf{Learner experience.}
Attitudes toward AI improved significantly after the task, but satisfaction was predicted by within-turn production fluency, not by uptake. Learners who paused more within their turns reported lower satisfaction, suggesting that the subjective quality of AI interaction hinges on whether the learner can produce fluent speech in the space the AI provides, not on how much language they absorb from the AI's input. 
% This dissociation is consistent with syntactic priming as an implicit process \citep{McDonough_2006}: learners may reuse recently available structures without consciously noticing that reuse as a benefit.

\section{Conclusions}
\label{sec:conclusions}
AI and human dialogue scaffold L2 production through complementary mechanisms: structural priming through regular input in one case, interactional competence through coordinative demand in the other. The design challenge is not to make AI dialogue more like human dialogue, but to exploit what each mode does well.

% AI and human dialogue scaffold L2 production through different mechanisms: structural priming through regular input in one case, interactional competence through coordinative demand in the other. These are complementary affordances, not competing ones. The core challenge for system design is not to make AI dialogue more like human dialogue, but to exploit what each mode does well.

This points to concrete principles for integrating voice-based AI into language curricula:
For curricula, this suggests four design principles. First, AI tasks are well suited to focused morphosyntactic practice, where regular input can support priming of constructions such as dative prepositions and gender agreement. Second, AI practice should complement rather than replace peer speaking, since human dialogue exercises turn competition, clarification requests, repair, and collaborative floor management. Third, AI verbosity should be calibrated: shorter, less complete AI turns may reduce floor dominance and create more space for learner-initiated language. Finally, uptake metrics can support adaptive feedback on vocabulary and constructions adopted, missed, or produced incorrectly from ASR transcripts.

\textbf{Limitations.} Several limitations apply. Uptake measures cannot fully separate priming from parallel activation driven by shared task context; causal claims require manipulating interlocutor language. NLP pipelines trained on Standard German may degrade on disfluent L2 speech; word-level ASR errors were not corrected, although German-speaking authors verified speaker labels, task boundaries, and broad coherence. First-use counting captures breadth rather than depth of uptake, and the lexical first-source-wins rule creates a task-order confound ($d = 1.71$), addressed with lower-powered position-matched comparisons. The study cannot determine whether short-term priming becomes durable learning; delayed post-tests are needed. Transcript-based analysis also excludes prosody, pronunciation, stress, intonation, and segmental accuracy. The comparison involved stimulus and modality asymmetries: human partners viewed images directly and interacted face-to-face or by Zoom, whereas the AI received English text descriptions and interacted through audio only. Some temporal effects may therefore partly reflect interface latency, VAD behavior, and visual coordination cues rather than interlocutor type alone. Finally, the AI condition reflects one system configuration (model, voice, prompt, VAD setting, and task interface) with university German learners, mostly Dutch- or English-dominant, performing one task type; future work should replicate across AI systems, L1 backgrounds, proficiency levels, target languages, and tasks.

\clearpage
\section*{Acknowledgments}
We would like to thank all of the professors, teachers and students who worked with great dedication to plan for, collect and process all of the data for these experiments.

\bibliographystyle{abbrv}
\bibliography{spotter_edm_2026_clean}

\balancecolumns

\end{document}